\documentclass[review]{elsarticle}

\usepackage{lineno,hyperref}
\usepackage{textgreek}
\usepackage{xspace}
\usepackage{siunitx}
\usepackage{amsmath}
\modulolinenumbers[5]

\journal{to be determined}

\newcommand{\potassium}{K\textsuperscript{+}\xspace}
\newcommand{\ip}{IP\textsubscript{3}\xspace}
\newcommand{\calcium}{Ca\textsuperscript{2+}\xspace}

\begin{document}

\begin{frontmatter}

\title{A control mechanism for intramural periarterial drainage via astrocytes: How neuronal activity could improve waste clearance from the brain.}

\author[ced]{Alexandra K. Diem\fnref{correspondingauthor}}
\cortext[mycorrespondingauthor]{Corresponding author}
\ead{A.K.Diem@soton.ac.uk}
\author[med]{Roxana O. Carare}
\author[ced]{Neil W. Bressloff}

\address[ced]{Computational Engineering and Design, Faculty of Engineering and the Environment, University of Southampton, Southampton, UK}
\address[med]{Clinical Neurosciences, Faculty of Medicine, University of Southampton, Southampton, UK}

\begin{abstract}
The mechanisms behind waste clearance from deep within the parenchyma of the brain remain unclear to this date. Experimental evidence has shown that one pathway for waste clearance, termed intramural periarterial drainage (IPAD), is the rapid drainage of interstitial fluid (ISF) via basement membranes (BM) of the smooth muscle cells (SMC) of cerebral arteries and its failure is closely associated with the pathology of Alzheimer's disease (AD) \cite{Carare2008,Arbel-Ornath2013}. We have previously shown that arterial pulsations from the heart beat are not strong enough to drive waste clearance \cite{Diem2017}. Here we demonstrate computational evidence for a mechanism for cerebral waste clearance that is driven by functional hyperaemia, that is, the dilation of cerebral arteries as a consequence of increased neuronal demand. This mechanism is based on our model for fluid flow through the vascular basement membrane \cite{Diem2017}. It accounts for waste clearance rates observed in mouse experiments and aligns with pathological observations as well as recommendations to lower the individual risk of AD, such as keeping mentally and physically active. 
\end{abstract}


\end{frontmatter}

To this date the driving mechanisms behind waste clearance from the brain, which lacks traditional lymphatics, remain largely unclear. However, clinical evidence suggests a pathway for waste clearance via the basement membranes (BM) of smooth muscle cells (SMC) within artery walls, termed intramural periarterial drainage (IPAD). Failure of IPAD is closely associated with the pathology of Alzheimer's disease (AD) and has been shown to fail as a consequence of a loss of the heart beat \cite{Carare2008,Arbel-Ornath2013,Morris2014}, leading to the  hypothesis that arterial pulsations due to the heart beat drive waste clearance from the brain \cite{Weller2009,Iliff2012,Asgari2015,Sharp2016}. However, our previous work demonstrates that arterial pulsations are not strong enough to drive IPAD and thus the exact mechanisms behind waste clearance from the brain remain unclear \cite{Diem2017}. Here, using computational simulations, we demonstrate how functional hyperaemia, which describes the dilation of cerebral arteries due to an increase in neuronal activitiy, may provide the driving mechanism for cerebral waste clearance. Chemical communication within the neurovascular unit (NVU), which comprises neurons, astrocytes and arteries \cite{Haydon2006}, can lead to a dilation of \SI{20}{\percent} in a small artery of around \SI{40}{\um} diameter \cite{Farr2011,Witthoft2012}. Combined with our model of flow through the BM \cite{Diem2017} and a finite element (FE) model of an artery wall we successfully demonstrate waste clearance at a rate comparable to experimental observations \cite{Carare2008}. Additionally we demonstrate how this mechanism is impaired by a stiffening of the artery wall, as commonly occurs during age and leads to risk factors for AD such as atherosclerosis \cite{Ostergaard2013}. Our mechanism is in line with key pathological features of AD and recommendations to reduce the individual risk of developing AD via physical and mental activity by the Alzheimer's Society \cite{AlzSoc2016}.

\begin{figure}
\centering
\includegraphics[width=\linewidth]{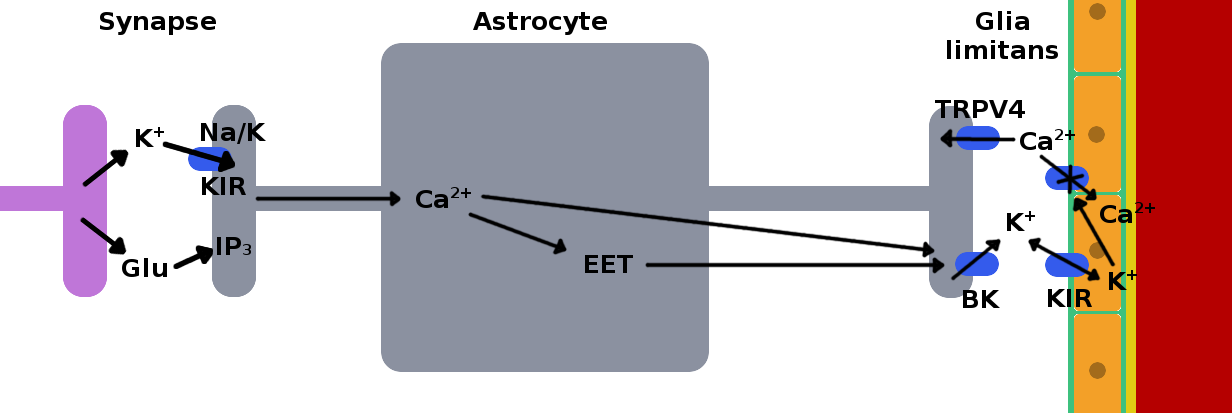}
\caption{The synapses of active neurons release glutamate (Glu) and potassium (\potassium). \potassium enters the astrocyte via Na-K pumps and inwardly rectifying \potassium (KIR) channels, whilst Glu receptors in the astrocyte lead to the production of inositol triphosphate (\ip). Internal calcium (\calcium) stores are released and lead to the production of epoxyeicosatrienoic acids (EET), which opens big \potassium (BK) channels at the astrocyte endfoot, releasing \potassium. KIR channels of SMC of the artery activate and increase the extracellular \potassium concentration, which closes \calcium channels, decreasing \calcium influx into the SMC and dilating the artery. The dilation activates stretch-gated transient receptor potential channels 4 (TRPV4) on the astrocyte endfoot, leading to an influx of extracellular \calcium into the astrocyte \cite{Haydon2006}.\label{fig:nvu}}
\end{figure}

\begin{figure}
\centering
\includegraphics[width=0.6\linewidth]{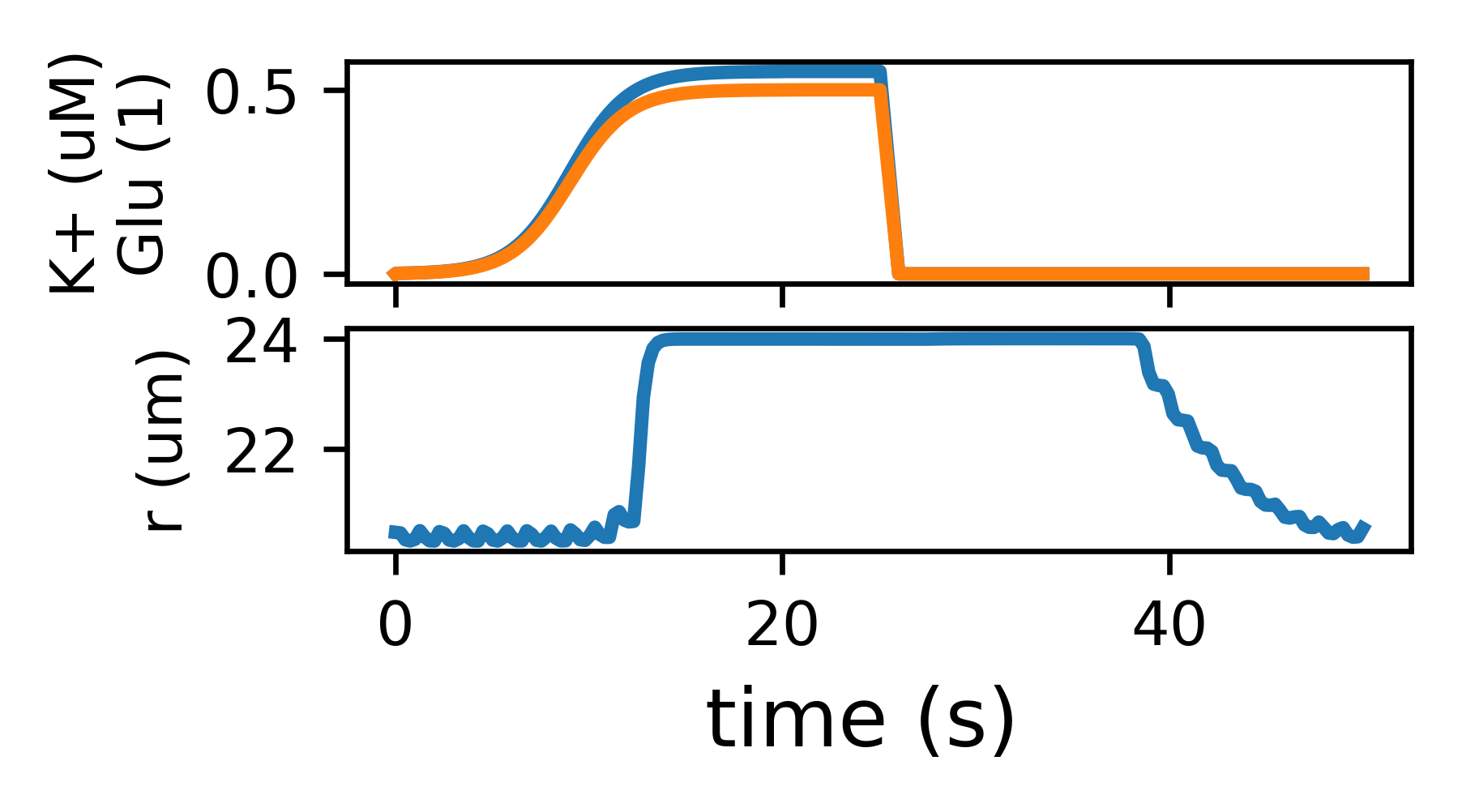}
\caption{Effect of the release of \potassium and Glu into the synaptic space of a neuron and an astrocyte process. Glu here refers to the ratio of bound/unbound Glu receptors (dimensionless). The arterial radius is modelled using a system of ODE \cite{Witthoft2012} with corrections to the equations listed in \cite{Diem2017a}. The model shows a dilation of the artery of \SI{20}{\percent}.\label{fig:nvu_model}}
\end{figure}

The brain is the fastest metabolising organ in the body and has unique energy demands. Whilst only taking up \SI{2}{\percent} of our body mass it consumes \SI{20}{\percent} of our energy \cite{Zlokovic2008}. Thus, adequate nutrient supply to neurons is vital in maintaining brain function. Functional hyperaemia is the mechanism by which the NVU increases blood flow to an active region of the brain (Figure~\ref{fig:nvu}). The synapses of active neurons release glutamate (Glu) and potassium (\potassium), which leads to the release of internal \calcium stores from the astrocyte via its endfeet. The SMC membrane is depolarised and the artery dilates \cite{Haydon2006}. This chemical signalling cascade has been modelled by \cite{Witthoft2012} using ordinary differential equations (ODE), and an open source implementation of the model is found in \cite{Diem2017a}. The dilation of an artery in response to a stimulus is shown in Figure~\ref{fig:nvu_model} and the displacement $U(t)$ of the artery is defined as
\begin{equation}
U(t) = r(t) - r_0,\\
\end{equation}
where $r_0$ is the arterial radius at rest.

We modelled and artery wall based on an arteriole with radius $r_0 = \SI{20}{\um}$ as a two-dimensional rectangle of height $h = \SI{4}{\um}$ and length $l = \SI{200}{\um}$ using the FE solver FEniCS \cite{Alnaes2015}, using a linear elastic model with the governing equations
\begin{gather}
  - \nabla \cdot \sigma = 0 \qquad \text{on } \Omega\\
  \sigma = \lambda (\nabla \cdot u) I + \mu (\nabla u + (\nabla u)^T),
\end{gather}
where $\Sigma$ describes the domain, $u$ is the artery wall displacement, $\sigma$ is the stress tensor, $I$ is the identity matrix and $\lambda = \SI{16.44}{\mega\pascal}$ and $\mu = \SI{335.57}{\kilo\pascal}$ are the Lam\'e coefficients. The radial boundaries were fixed in space while the longitudinal boundaries were allowed to deform freely. Additionally, an astrocyte was placed centrally on the outer longitudinal boundary, applying the Dirichlet boundary condition
\begin{equation}\label{ast_bnd}
u = U(t) \qquad \text{on } \partial \Omega_A,\\
\end{equation}
where $\Omega_A$ refers to a section of the outer longitudinal boundary, which models an astrocyte. The optimal mesh consisted of 33,348 triangular elements. The governing equations were solved for \SI{50}{\s} with a time step of \SI{0.25}{s}, where at each time step the boundary condition (\ref{ast_bnd}) was updated.

The rate of IPAD $q = (q_1, q_2)$ was calculated using our model based on Darcy's law for porous media under thin-film flow conditions \cite{Diem2017} in cylindrical coordinates is governed by
\begin{gather}
  q = q_1 \boldsymbol{e_z} + q_2 \boldsymbol{e_r} = - K(\frac{\partial p}{\partial z}) (\frac{\partial p}{\partial z} \boldsymbol{e_z} + \frac{\partial p}{\partial r} \boldsymbol{e_r})\\
  \frac{\partial q_1}{\partial z} + \frac{1}{r} \frac{\partial}{\partial r} \left( r q_2 \right) = 0,
\end{gather}
where $\boldsymbol{e_z}$ and $\boldsymbol{e_r}$ describe the unit vectors in the $z$ and $r$ direction and $K(\partial p/ \partial z)$ describes the permeability of the BM. Permeability is modelled as a step function
\begin{equation}
  K(\frac{\partial p}{\partial z}) = \frac{k}{\mu} \cdot \begin{cases}
    K_0 & \text{if } \partial p/\partial z < 0,\\
    K_1 & \text{otherwise,}
  \end{cases}
\end{equation}
where $k = \SI{1e-12}{\square\cm}$ represents the intrinsic permeability of the extracellular matrix, $\mu = \SI{1.5e-3}{\pascal\s}$ represents the viscosity of ISF and the ratio $0 < K_0/K_1 \leq 1.0$ determines the strength of the valve \cite{Diem2017}. Applying thin-film flow conditions and kinematic boundary conditions the model equation
\begin{equation}
  \frac{\partial}{\partial t} \left( \gamma \cdot R_i(z, t) \cdot h(z,t) \right) = \frac{\partial}{\partial z} \left( R_i(z, t) \cdot h(z,t) \cdot K(\frac{\partial p}{\partial z}) \frac{\partial p}{\partial z} \right) 
\end{equation}
is derived (see \cite{Diem2017} for details). The IPAD model requires the pressure gradient along the artery $\partial p / \partial z$, which can be recovered directly from the stresses inside the artery wall, such that
\begin{equation}
p = - \sigma_{rr},
\end{equation}
where $\sigma_{rr}$ describes the principal component of $\sigma$ in the $r$-direction \cite{Diem2016}.

\begin{figure}
\centering
\includegraphics[width=\linewidth]{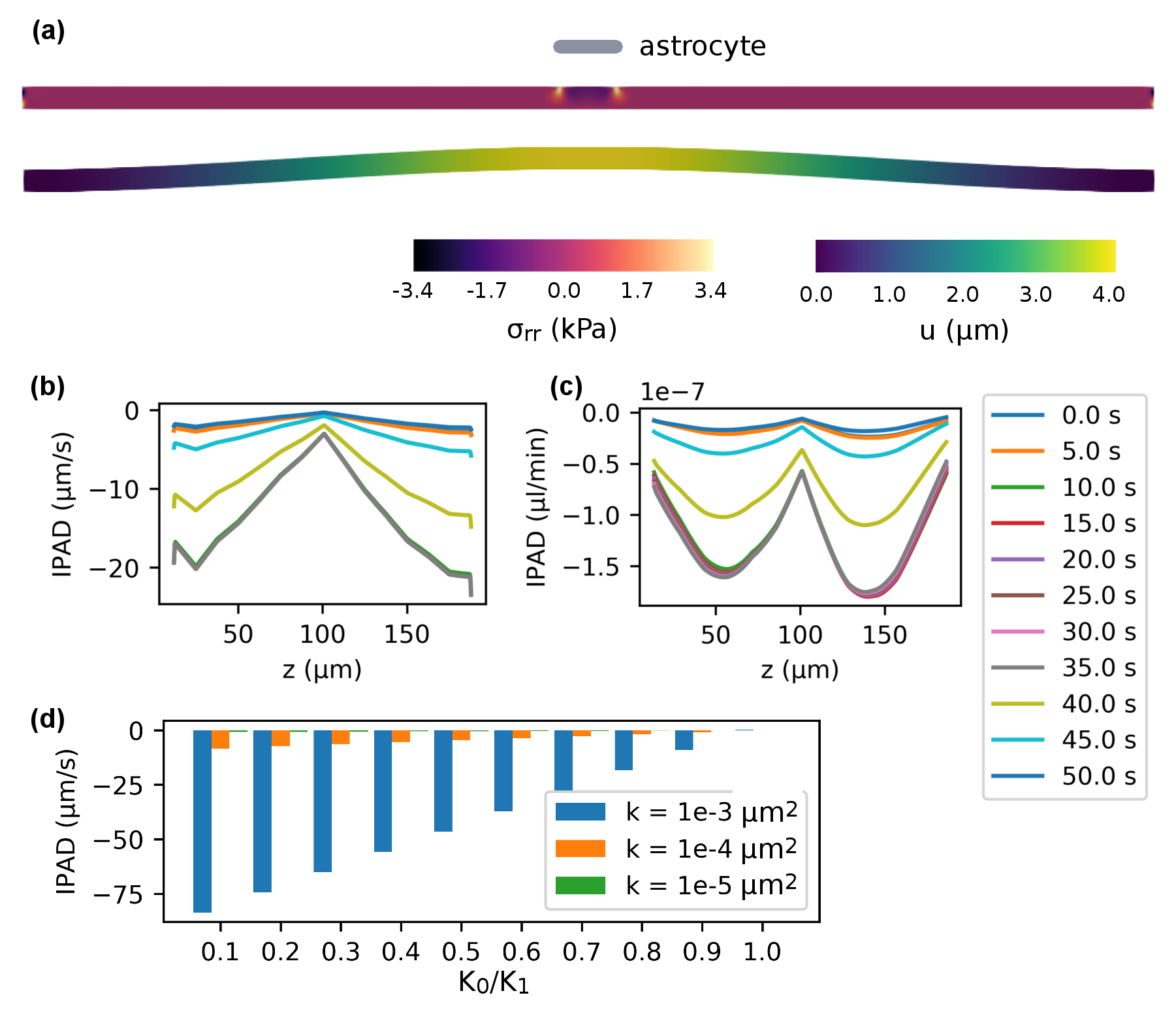}
\caption{IPAD inside a cerebral artery. (a) Displacement and stress of the artery wall due to $U(t)$ at $t = \SI{20}{\s}$ of a single astrocyte end-foot. Because displacement is fixed to $u = 0$ at the ends, stresses at the ends are high. Thus, all following results are presented for $\SI{10}{\um} \leq z \leq \SI{190}{\um}$. (b) IPAD velocity at various time points over the length of the artery wall using $K_0/K_1 = 0.1$. The average velocity over time and space is \SI{-8.37}{\um\per\s}. (c) IPAD flow rate at various time points over the length of the artery wall using $K_0/K_1 = 0.1$. The average flow rate over time and space is \SI{-7.24e-8}{\ul\per\min} for a single arteriole. Extrapolated over 6.5 billion arterioles estimated for the human brain it would take \SI{9.92}{\hour} to process the total amount of ISF in the brain (\SI{280}{\ml}). (d) IPAD velocity for various values of ECS permeability $k$ over the strength of the valve mechanisms $K_0/K_1$. Values are always negative, except for when $K_0/K_1 = 1.0$. The effect of the valve mechanism is decreased with decreasing $k$.\label{fig:results}}
\end{figure}

The results are shown in Fig.~\ref{fig:results}. Displacement and stress inside the artery wall are shown (Fig.~\ref{fig:results}a) for a single astrocyte of length \SI{10}{\um} at $t = \SI{20}{\s}$, where dilation maximal dilation of the artery has been achieved (see Fig.~\ref{fig:nvu_model}). The stress plot shows high stresses at both ends of the artery, which are due to the wall being fixed to $u = 0$. Thus, to avoid an influence of these boundary effects on the results all further results are presented for $\SI{10}{\um} \leq z \leq \SI{190}{\um}$. Experimental results from \cite{Carare2008} suggest an expected IPAD velocity of \SI{8}{\um\per\s} for the mouse brain, which is adopted as a benchmark value here. Using $k = \SI{1e-4}{\square\um}$ and $K_0/K_1 = 0.1$ an average IPAD velocity of \SI{-8.37}{\um\per\s} is achieved, where the negative sign indicates flow in the reverse direction of the blood flow as desired (Fig.~\ref{fig:results}b). The average flow rate for the same parameters is \SI{-7.24e-8}{\ul\per\min} for a single arteriole (Fig.~\ref{fig:results}c). The total length of capillaries in the brain is 400~miles \cite{Nelson2016}. Estimating the length of an average capillary as \SI{100}{\um} the number of capillaries in the brain is 6.5 billion. Thus, it would take approximately \SI{9.92}{\hour} to process the full volume of ISF (\SI{280}{\ml}). In comparison CSF has a turnover rate of three to four times a day \cite{Hladky2014}.

Permeability $k$ of the extracellular space in the brain is not known, but it ranges between \SI{0.01}{\square\um} and \SI{1e-6}{\square\um} for interstitial spaces elsewhere in the body \cite{Levick1987}. Thus it is interesting to observe IPAD velocities for different values of $k$ and valve strength (Fig.~\ref{fig:results}d). The target velocity of $\approx \SI{8}{\um\per\s}$ was achieved for $k = \SI{1e-4}{\square\um}$, and net velocity is always negative except for $K_0/K_1 = 1.0$, i.~e. absence of a valve mechanism. A change of the permeability by one order of magnitude results in the same change in the IPAD velocity, thus the effect of the valve mechanism decreases with decreasing $k$.

\begin{figure}
\centering
\includegraphics[width=\linewidth]{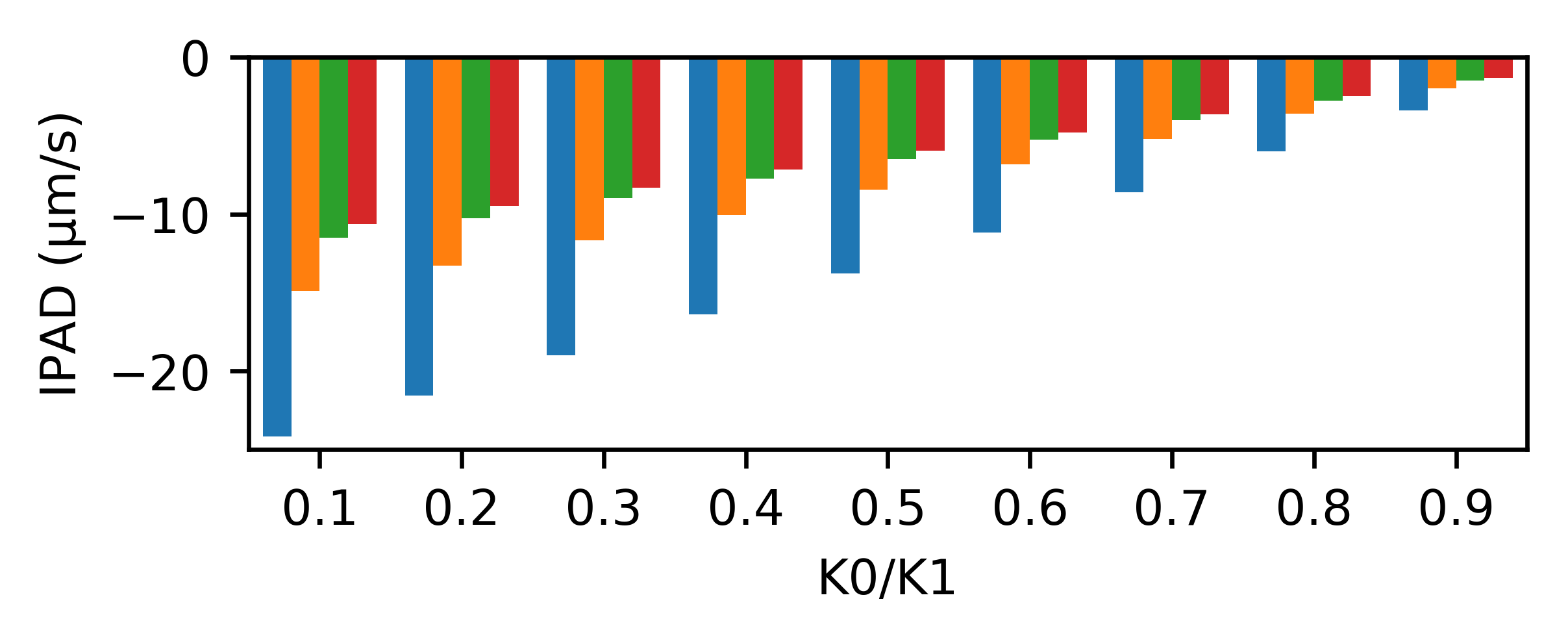}
\caption{Comparison of IPAD velocities on an arteriole for varying numbers of astrocytes acting on the wall and valve strength. An astrocyte end-foot has a width of \SI{10}{\um} and the gap between astrocytes is \SI{1}{\um}. Length has a negative effect on IPAD whilst the number of astrocytes acting on the arteriole simultaneously has a positive effect. $k = \SI{1e-3}{\square\um}$, arteriole length $l = \SI{309}{\um}$, number of astrocytes: 10 (blue), 5 (orange), 2 (green), 1 (red). \label{fig:vel_lengths}}
\end{figure}

Fig.~\ref{fig:vel_lengths} shows a comparison of IPAD for varying numbers of astrocytes on a longer arteriole with $l = \SI{309}{\um}$. Astrocyte end-feet are estimated to be \SI{10}{\um} in width with \SI{1}{\um} gaps in between. Increasing the arteriole length leads to a decrease in IPAD, whilst maintaining a suitable velocity. Increasing the number of astrocytes increases IPAD. It is interesting to note that IPAD is not neccessarily positive at $K_0/K_1 = 1.0$, however, if it remains negative its absolute value is more than an order of magnitude smaller compared to $K_0/K_1 = 0.1$, indicating stagnating flow. The dilation of the arteriole here is not caused by a travelling wave and hence directionality of ISF flow may not be dictated by the dilation on its own. Removing the gaps between astrocytes leads to a reduction of IPAD, which, for ten astrocytes at $K_0/K_1 = 0.1$ reduces from \SI{-24.16}{\um\per\s} to \SI{-22.06}{\um\per\s}.

Whilst our model of flow through the BM has previously disproved the well-cited hypothesis that arterial pulsations due to the heart rate drive IPAD \cite{Diem2017}, we have used it here to demonstrate a more realistic driving mechanism utilising the NVU. All parameters chosen lie well within physiologically relevant ranges found in the literature. The mechanism produces velocities in accordance with experimental results from \cite{Carare2008} and the total estimated flow rate suggests a turnover of ISF just under three times a day, similar to the turnover rate for CSF.

Our study supports the new hypothesis that arterial dilations due to neurovascular coupling may play a crucial role in waste clearance from the brain. This hypothesis is more intuitive than the previously most cited hypothesis of pulsations due to the heart beat driving waste clearance: Athletes have a comparatively low heart rate, which would put them at increased risk of developing AD under the heart beat hypothesis. However, research shows that regular physical activity acts as a preventive mechanism \cite{Woodard2012} and is generally beneficial for cognitive ability \cite{Gomez-Pinilla2013}. It decreases amyloid load in transgenic mouse models \cite{Adlard2005} and reduces the risk of hippocampal atrophy in individuals at genetic risk of developing AD \cite{Smith2014}. Our neurovascular coupling hypothesis supports and can explain these findings. Physical activity leads to increased neuronal activity, which demands better nutrient supply and, according to our model, is accompanied by increased IPAD and thus decreased amyloid load. Analogously, the same recommendation applies more generally for any cognitively engaging type of activity \cite{Scarmeas2001,Wilson2002,Fratiglioni2004} and is equally well explained by our hypothesis and supported by our model. In addition our hypothesis supports the findings that cardiovascular diseases such as atherosclerosis and hypertension pose risk factors for AD \cite{Ostergaard2013}, most likely via two mechanisms: a decresed response to the astrocyte due to a stiffening of the artery wall and stagnating IPAD due to a dysfunction of the valve mechanism. We conclude that our neurovascular hypothesis of the mechanism behind IPAD offers suitable explanations for cardiovascular findings associated with AD and should thus act as the new working hypothesis for the mechanism behind IPAD.

\section*{Acknowledgements}

This work was supported by an EPSRC Doctoral Training Partnership grant (EP/N509747/1).

\bibliography{nvu-drainage}

\end{document}